\documentclass[12pt]{article}
\usepackage{amsfonts}
\usepackage{amsmath,amssymb}
\usepackage{bbm}
\usepackage[latin1]{inputenc}
\usepackage{fancyhdr}

\def\Dmm{\mathrm{D}^{--}}
\def\Dpp{\mathrm{D}^{++}}
\def\epsilonm{\epsilon^-{}}

\def\ppp{\partial^{++}}
\def\Psibm{\bar{\Psi}^-{}}

\def\thetabp{\bar{\theta}^+{}}
\def\thetam{\theta^-{}}
\def\thetap{\theta^+{}}
\def\Vmm{V^{--}}

\def\V{V^{++}_{\text{WZ}}}

\def\phib{\bar{\phi}}
\def\Psib{\bar{\Psi}}

\def\thetab{\bar{\theta}}

\def\bPsi{\bar{\Psi}}
\def\phib{\bar{\phi}}

\def\bg{\bar{g}}

\def\ve{\varepsilon}

\def\alphad{{\dot{\alpha}}}

\def\lambdad{\dot{\lambda}}

\def\dalpha{\dot{\alpha}}

\def\Ac{{\cal A}}
\def\Wc{{\cal W}}

\newcommand{\be}{\begin{equation}}
\newcommand{\ee}{\end{equation}}
\newcommand{\bea}{\begin{eqnarray}}
\newcommand{\eea}{\end{eqnarray}}
\newcommand{\nn}{\nonumber\\ }
\newcommand{\pe}[1]{(\ref{#1})}

\newcommand{\lb}{\label}

\def\iim{\mathrm{i}}

\def\L{\Lambda_\epsilon}

\title{Non-singlet Q-deformations of ${\cal N}=2$  gauge theories \footnote{To be published in
the {\it Proceedings of the International Workshop "Supersymmetries and Quantum
 Symmetries"}, JINR,  Dubna, 2005}}

\author{A. De Castro$\,{}^a$, E. Ivanov$\,{}^b$,
        O. Lechtenfeld$\,{}^a$ and L. Quevedo$\,{}^a$}

\date\empty

\begin{document}

\maketitle

\begin{center}

$^a$ Institut f\"ur Theoretische Physik, Universit\"at Hannover, \\
Appelstrasse 2, 30167 Hannover, Germany. \\
{\tt castro, lechtenf, quevedo@itp.uni-hannover.de}

\vspace{3mm}

$^b$ Bogoliubov Laboratory of Theoretical Physics, JINR, \\
141980 Dubna, Moscow Region, Russia.\\
{\tt eivanov@theor.jinr.ru}
\end{center}
\vspace{5mm}
\begin{abstract}
\noindent
We study a non-anticommutative chiral non-singlet deformation of the
${\cal N}{=}(1,1)$ abelian gauge multiplet in Euclidean harmonic
superspace.  We present a closed form of the gauge transformations and
the unbroken ${\cal N}{=}(1,0)$ supersymmetry transformations preserving
the Wess-Zumino gauge, as well as the bosonic sector of the ${\cal
N}{=}(1,0)$ invariant action. This contribution is a summary of our main
results in \texttt{hep-th/0510013}.
\end{abstract}
\vspace{5mm}
Extensions of gauge theories to non-commutative and
non-(anti)commutative superspaces are currently of remarkable interest
within the high energy physics community, mainly due to their relevance
to subjects like string theory (see for example
\cite{DouglasdeBoerOoguri, Seiberg:2003yz}  and references therein).
Here we focus in a subclass of non-(anti)commutative Euclidean
supersymmetric field theories called \emph{Q-deformed}, realized via a
Weyl-Moyal product with a bilinear nilpotent Poisson operator, which is
constructed  in terms of the supercharges,
\begin{equation}\lb{Poisson}
P = -\overleftarrow{Q}^i_\alpha C^{\alpha \beta}_{ik}
\overrightarrow{Q}^k_\beta\,.
\end{equation}
The Moyal product of two superfields is then defined by 
\begin{equation}\label{Mp}
A\star B=A e^P B\,. 
\end{equation}
The deformation parameters $C^{\alpha\beta}_{ij}$ form a constant tensor
which is symmetric under the simultaneous permutation of the Latin and
Greek indices, ${C}^{\alpha\beta}_{ij}={C}^{\beta\alpha}_{ji}$.
Generically, it breaks the full automorphism symmetry $Spin(4)\times$
O(1,1)$\times$ SU(2) $\equiv$ SU(2)$_{\text{L}}\times$
SU(2)$_{\text{R}}\times$ O(1,1)$\times$ SU(2) of the ${\cal N}=(1,1)$
superalgebra ---O(1,1) and SU(2) being the R-symmetry groups--- down to
SU(2)$_\text{R}$.  An important feature of Q-deformations is the nilpotent
nature of the Poisson operator ($P^5=0$) which  makes the Moyal product
polynomial, ensuring local actions. In virtue of the commutation
properties of $P$ with respect to the spinor covriant derivatives,
\begin{equation}\label{pconmutacond}
\left[D^{\pm}_\alpha , P\right] = 0\,, \qquad
\left[\bar{D}^{\pm}_{\alphad} , P\right] = 0\,, \qquad
\left[D^{\pm\pm} , P\right] = 0,
\end{equation}
the product \eqref{Mp} breaks ${\cal N}=(1,1)$ supersymmetry down to
${\cal N}=(1,0)$ while preserving both chirality and Grassmann harmonic
analyticity of the involved superfields, as well as the harmonic
conditions\footnote{Giving up chirality and analiticity, it is also
possible to use spinor covariant derivatives to construct a nilpotent
Poisson operator. We recommend ref.\cite{KetovKlemmFerraraLledoTamassia}
for a deeper treatment of the subject.}$D^{\pm\pm} A=0$. Operator
\eqref{Poisson} can be split as follows,
\begin{equation}
P = -I \overleftarrow{Q}^i_\alpha \ve^{\alpha \beta}\ve_{ik}
\overrightarrow{Q}^k_\beta -\overleftarrow{Q}^i_\alpha
\hat{C}^{\alpha \beta}_{ik}
\overrightarrow{Q}^k_\beta\,. \label{Generic}
\end{equation}
The first term is $Spin(4)\times$SU(2)-preserving while the second term
involves a SU(2)$_\text{L}\times$SU(2) constant tensor which is
symmetric under the independent permutations of Latin and Greek indices,
$\hat{C}^{\alpha\beta}_{ij}=
\hat{C}^{\beta\alpha}_{ij}=\hat{C}^{\alpha\beta}_{ji}$. For the generic
choice, it fully breaks Euclidean symmetry, SU(2)$_{\text{L}}$ and
R-symmetry SU(2).  Q-deformations induced only by the first term are
called \emph{singlet} or \emph{QS-deformations}, whereas those
associated with the second term, \emph{non-singlet} or
\emph{QNS-deformations}.  In this contribution we report important
results on dynamical aspects of \emph{QNS-deformations} of the $N=(1,1)$
U(1) vector multiplet in harmonic superspace. The talk is based on paper
\cite{DeCastro:2005vb}, where detailed calculations are performed and a
complete list of references is given.
\paragraph{Gauge transformations.}
The residual gauge transformations of the component fields of the
Abelian ${\cal{N}}=(1,1)$ vector multiplet in the WZ gauge can be found
from the Q-deformed superfield transformation \cite{Ferrara:2004zv}
\begin{equation}
\delta_\Lambda \V = \Dpp\Lambda+[\V, \Lambda]_{\star}\,, \label{DeltaV}
\end{equation}
with $\V$ being the analytic harmonic U(1) superfield gauge connection
and $\Lambda$ the analytic residual gauge parameter satisfying
$D^{+}_\alpha \Lambda = \bar{D}^{+}_{\alphad} \Lambda = 0\,$.  In the
left-chiral basis, where $x_A^{\alpha\alphad}= x_L^{\alpha\alphad}
-4\iim\theta^{-\alpha}\thetab^{+\alphad}$ \cite{Galperin:2001uw}, $\V$
has the following $\theta$-expansion
\begin{equation}
\begin{aligned}
\V & = (\theta^+)^2 \phib+\thetab^+_{\alphad}\left[ 
2\theta^{+\alpha}A_{\alpha}^{\alphad} +
4(\theta^+)^2\Psib^{-\alphad}
-2\iim (\theta^+)^2\theta^{-\alpha}\partial_\alpha^{\alphad}\,\phib\right]\\&
+(\thetab^+)^2\Bigl[\phi + 4\theta^+\Psi^- + 3(\theta^+)^2 D^{--}
-\iim (\theta^+\theta^-)\partial^{\alpha\alphad}A_{\alpha\alphad}
+ \theta^{-\alpha}\theta^{+\beta}\,F_{\alpha\beta}\\ &
-(\theta^+)^2(\theta^-)^2\square\phib+4\iim\,
(\theta^+)^2\theta^{-\alpha}\partial_{\alpha\dalpha}\bPsi^{-\dalpha}\Bigr].
\end{aligned}
\end{equation}
The superparameter $\Lambda_0= \iim a +2\theta^{-\alpha}
\bar\theta^{+\alphad} \partial_{\alpha\alphad}a -\iim(\thetam)^2
(\thetabp)^2 \Box\, a\,$ (being $a$ an arbitrary function of $x_L$) found
for the undeformed and singlet cases, breaks the WZ gauge in the
non-singlet case, due to the appearance of an unwanted dependence on the
harmonic variables $u^{\pm}_i$ in the expresions for the gauge
variations. It is clear that is imperative to choose a gauge parameter
$\Lambda$ that preserves the WZ gauge, that is, for non-singlet deformations,
some correction terms $\Delta\Lambda $ must be added to $\Lambda_0$, where
\begin{eqnarray}
&& \Delta\Lambda = \theta^{+}_\alpha \bar\theta^+_{\alphad}\,
\partial^{\alphad}_\beta\,a\,B^{--\alpha\beta}_1
+ (\bar\theta^+)^2 \partial_{\beta\dot\beta}\,a\,
A^{\dot\beta}_\alpha G^{--\alpha\beta} +
(\theta^+)^2(\bar\theta^+)^2 \,\Box a P^{-4} \nn
&& +\, (\bar\theta^+)^2\theta^+_\alpha \,\left[\,\Psib^{-\dot\beta}
\partial_{\beta\dot\beta}\,a\,B_2^{--\alpha\beta} +
\Psib^{+\dot\beta}\partial_{\beta\dot\beta}\,a\,G^{-4\alpha\beta}\,\right]
+(\theta^+)^2(\bar\theta^+)^2\, \partial_{\alpha\alphad}a\,
\partial^{\alphad}_\beta\,\phib\,B^{-4\alpha\beta}_3 \nn
&& +\iim\,\theta^+_\alpha\theta^-_\beta\, (\bar\theta^+)^2\,\Box\,a
\,B^{--\alpha\beta}_1 + \iim\, \theta^+_\alpha \theta^-_\gamma
(\bar\theta^+)^2\,\partial_{\beta\dot\lambda}a\,
\partial^{\gamma\lambdad}\phib\,\frac{d}{d\phib}B^{--\alpha\beta}_1\,.
\label{25}
\end{eqnarray}
The coefficients in \eqref{25} are some undetermined functions of
harmonics, the field $\phib$ and deformation parameters, calculated by requiring
\begin{equation}
\partial^{++}\delta A_{\alpha\alphad} = 0,\quad
\partial^{++}\delta \phi = 0,\quad
(\partial^{++})^2\delta \Psi^-_\alpha
= 0,\quad(\partial^{++})^3\delta D^{--} = 0\, .
\end{equation}
The correction term to $\delta_0 \V$ is,
\begin{equation}
\hat{\delta}\V = \Dpp\Delta \Lambda + [\V, \Delta\Lambda]_\star\,,\label{26}
\end{equation}
and the full WZ preserving gauge transformations are given by $\delta \V
=\delta_0 \V + \hat{\delta} \V $.  Unfortunately, it is very difficult
to find closed solutions of these equations for general deformation
parameters, though their perturbative solutions always exist as series
expansion. For the general choice of
$\smash{\hat{C}^{ij}_{\alpha\beta}}$, the gauge and susy transformations
and the corresponding action are known only to few first orders in the
parameters of deformation \cite{Araki}. Nevertheless, exact solutions
can be found for the product structure
\begin{equation*}
\smash{\hat{C}^{ij}_{\alpha\beta}=b^{ij}c_{\alpha\beta}}\, ,
\end{equation*}
which correspons to the maximally symmetric non-singlet deformations. 
The full set of non trivial QNS-deformed
gauge transformations laws for the ${\cal{N}}=(1,1)$ vector multiplet in
WZ gauge are then

\begin{align}\label{gtoriginal}
\delta\,\phib= &0\,, \qquad \delta\Psib^k_{\alphad}=0\,, \\
\delta\,A_{\alpha\alphad}=& X\coth X \partial_{\alpha\alphad}a\, ,\\
\delta\,\phi= &2\sqrt{c^2\,b^2}\left( \frac{1-X\coth X}{X}\right)
A^{\alpha\alphad}\partial_{\alpha\alphad}a\, ,\\\nonumber
\delta\Psi^i_\alpha=&\Biggl\lbrace
\left[\frac{4X^2(X\coth X-1)}{X^2+\sinh^2X-X\sinh2X}
\right] b^{ij}c_{\alpha\beta}\\
&-\sqrt{c^2b^2}\,
\left[\frac{4X\cosh^2X-2X^2(\coth X+X)-\sinh2X }
{X^2+\sinh^2X-X\sinh2X}\right]\ve^{ij}\ve_{\alpha\beta}
\Biggr\rbrace \Psib_{j\alphad}\,\partial^{\beta\alphad}a\,.\\
\delta D_{ij}=& 2\iim
b_{ij}c^{\alpha\beta}\partial_{\alpha\alphad}\phib\,
\partial^{\alphad}_\beta a\,.
\end{align}
where 
\begin{equation}
X=2\phib\sqrt{b^{ij}b_{ij}\,c^{\alpha\beta}_{\alpha\beta}}.
\end{equation}
Detailed calculations of these
transformations laws are carried out in \cite{DeCastro:2005vb}.  Having
the explicit QNS-deformed gauge transformations, one can deduce
a minimal Seiberg-Witten-like map  which puts these transformations into
the standard undeformed form
\begin{align}
&\Psi^i_{\alpha}=\widetilde{\Psi}^i_{\alpha}+
2\sqrt{c^2b^2}
\left[2\left(\coth X - \frac{1}{X}\right) - X\right]
\Psib^{i\alphad}\widetilde{A}_{\alpha\alphad}\,, \quad
D_{ij}= \widetilde{D}_{ij}+2\iim
b_{ij}c^{\alpha\beta}\partial_{\alpha\alphad}\phib\,
\widetilde{A}^{\alphad}_\beta\,,\nonumber\\
& A_{\alpha\alphad}=\,\widetilde{A}_{\alpha\alphad}\, X\coth X , \quad
\phi=\,\widetilde{\phi}
+\widetilde{A}^2\;\sqrt{c^2\,b^2}\;X\coth X
\left( \frac{1-X\coth X}{X}\right) .\label{SW}
\end{align}
For the fields with tilde we obtain the standard transformations
\begin{equation*}
\begin{aligned}\label{gtcan}
&\delta \widetilde{A}_{\alpha\alphad}=
\partial_{{\alpha\alphad}}a\,,
\qquad \delta \widetilde\phi=0\,,\qquad
\delta\widetilde{D}^{ij}=0\,,\qquad
\delta\widetilde{\Psi}^{k}_{\alpha}=0\,.
\end{aligned}
\end{equation*}
The gauge field strength
$\smash{F_{\alpha\beta} = 2\iim\partial_{(\alpha\alphad}A^{\alphad}_{\beta)}}$
which is non-covariant with respect to the deformed transformations is
redefined under the transformation $\smash{A_{\alpha\alphad} \rightarrow
\widetilde{A}_{\alpha\alphad}}$ as
\begin{equation*}
F_{\alpha\beta}=\widetilde{F}_{\alpha\beta}
X\coth X+4\iim\sqrt{b^2\,c^2}\widetilde{A}_{(\beta\alphad}\partial^{\alphad}_{\alpha)}\phib
\left(\coth X - \frac{X}{\sinh^2{X}}\right), \label{FF}
\quad\text{where}\quad 
\widetilde{F}_{\alpha\beta}=2\iim\partial_{(\alpha\alphad}
\widetilde{A}^{\alphad}_{\beta)}.
\end{equation*}
\paragraph{Unbroken susy transformations.}
Unbroken supersymmetry is realized on $\V$ as
\begin{equation}\label{unbrokensusy}
\delta\V=\left( \epsilon^{+\alpha}\partial_{+\alpha}
+\epsilon^{-\alpha}\partial_{-\alpha}\right)\V
-\Dpp\Lambda_c-\left[ \V\, ,\,\Lambda_c \right]_\star\,,
\end{equation}
where the star bracket, like in the previous consideration, is defined
via the non-singlet Poisson structure with the deformation matrix
$\hat{C}^{ij}_{\alpha\beta}=b^{ij}c_{\alpha\beta}$ and $\Lambda_{c}$ is
the compensating gauge parameter which is necessary for preserving WZ
gauge.  As in the QNS-deformed gauge transformations case,
variations obtained using the original $\L\,$ for undeformed  and
singlet cases (see\cite{Ferrara:2004zv}), violate the WZ gauge due to an
unbalanced apperance of harmonic and Grassmann variables
\cite{DeCastro:2005vb}, so one is led to properly modify $\L\,$.  Thus
we define
\begin{equation}
	\Lambda_c = \L + F_\epsilon
\end{equation}
We denote by $\check\delta\V$ the lowest-order non-singlet part of the
transformations coming from the star commutator in \pe{unbrokensusy}
using $\L$, and rewrite \eqref{unbrokensusy} in the following way
\begin{equation}\label{unbrokensusy2}
\delta\V= \,\check\delta\V-\Dpp F_\epsilon
-\left[ \V\, ,\,F_\epsilon \right]_\star 
\end{equation}
with
\begin{equation}
    \begin{aligned}
    \L&=2(\epsilonm\thetap)\phib+
   \thetab^-_{\alphad}\left[
	4\iim(\epsilonm\thetap)\theta^-_\alpha\partial^{\alpha\alphad}\phib
    -2\epsilon^-_\alpha A^{\alpha\alphad}+
    4(\epsilonm\thetap)\Psibm^{\alphad}\right] +\\
    &(\thetabp)^2 \Bigl[2(\epsilon^-\Psi^-)
    +2\iim\epsilonm^{\alpha}\thetam^{\beta}
    \partial^{\alphad}_{\beta}A_{\alpha\alphad}
    -2(\epsilonm\thetap)(\thetam)^2\square\phib\\
    &\quad+4\iim(\epsilonm\thetap)\thetam^{\alpha}
    \partial_{\alpha\dalpha}\Psibm^{\dalpha}
    +2(\epsilonm\thetap)D^{--}\Bigr]\,.
    \end{aligned}
\end{equation}
The additional compensating gauge parameter intended
for restoring the WZ gauge with the minimal set of terms needed to eliminate the
improper harmonic and Grassmann dependence amounts to the following form
\begin{equation}
    \begin{aligned}
    F_\epsilon&=\theta^{+\alpha} \,f^-_\alpha
    +\thetab^+_{\alphad}\left[ {\bg}^{-\dalpha}
    +2\iim \theta^-_{\alpha}\,\theta^{+\beta}
    \partial^{\alpha\dalpha}\,f^-_{\beta}
    + \theta^{+\alpha}\,b^{--\dalpha}_\alpha
    +(\theta^+)^2\,{\bg}^{(-3)\dalpha}\right] \,\\
    +&(\thetab^+)^2\Bigl[g^{--} -(\theta^-)^2\theta^{+\alpha}\square f^-_\alpha
    +\iim\theta^{-\alpha}\partial_{\alpha\dalpha}{\bg}^{-\dalpha}
    +\iim \thetap^\alpha \thetam^\beta\partial_\beta^\alphad
    b^{--}_{\alpha\alphad}
    +\theta^{+\alpha}f^{(-3)}_\alpha\\&
    +\iim(\theta^+)^2\theta^{-\alpha}
    \partial_{\alpha\dalpha}{\bg}^{(-3)\dalpha}
    +(\theta^+)^2\,X^{(-4)}\Bigr]\,.
\end{aligned}
\end{equation}
Requiring the elimination of terms with unbalanced Grassmann variables and 
\begin{equation}
\ppp\delta\phib=0\,,\;
\left(\ppp\right)^2 \delta\bPsi^{-}_{\dalpha}=0\,,\;
\ppp\delta A_{\alpha\alphad}=0\,, \; (\partial^{++})^2\delta \Psi^{-}_\alpha = 0\,,
\; (\partial^{++})^3\delta D^{--} = 0\,,
\end{equation}
we can explicitly find  components of $F_\epsilon$ and restore the
correct ${\cal N}=(1,0)$ supersymmetry transformations preserving WZ
gauge. The full set of these transformations together with the full
supersymmetric action will be given\footnote{In fact, it is of no actual
necessity to explicitly know these transformations, since our procedure
of deriving the action is manifestly ${\cal N}=(1,0)$ supersymmetric by
construction \cite{DeCastro:2005vb}.} in \cite{In prep}. Here we show the
simplest subalgebra
\begin{equation}
\begin{aligned}
    \delta\phib&=0\,,\quad \delta A_{\alpha\dalpha}=
8\phib\epsilon^{i\beta}\bPsi^j_{\dalpha}\,b_{ij}c_{\alpha\beta}
+2\epsilon^{k}_{\alpha}\bPsi_{k\dalpha}\,
X\coth{X}\,, \\
    \delta\Psib^i_\alphad&=\left[
    \frac{2\iim}{\sqrt{b^2c^2}}\cosh X\,\sinh X\,
    c^{\alpha\beta}b^{ij}
    -\iim \cosh^2 X\,\ve^{\alpha\beta}\ve^{ij}\right]
    \epsilon_{j\beta}\partial_{\alpha\alphad}\phib\,.
\end{aligned}
\end{equation}
These variations form an algebra which is closed modulo a gauge
transformation with the composite parameter $a_c =
-2\iim(\epsilon\cdot\eta)\bar\phi\,$:
\begin{equation*}
\left[\delta_\epsilon,\delta_\eta \right]\phib=0\,,\quad
\left[\delta_\epsilon,\delta_\eta \right]\bPsi^j_{\dalpha}=0\,,\quad
\left[\delta_\epsilon,\delta_\eta \right]A_{\alpha\dalpha}=
-2\iim(\epsilon\cdot\eta)
\left(X\coth X \right)\partial_{\alpha\dalpha}\phib\,.
\end{equation*}
\paragraph{Bosonic action.}
Now we present the bosonic sector of the ${\cal{N}}=(1,0)$ gauge
invariant action in components.  The QNS-deformed action  for the
${\cal{N}}=(1,1)$ U(1) gauge theory in harmonic superspace
\cite{Galperin:2001uw}, in the form most appropriate for our purposes,
is written in the same way as in the QS-deformed case
\cite{Ferrara:2004zv}
\begin{equation}\label{actionWc}
    S=\frac14\int d^4x_L\,d^4\theta\,du\,\Wc\star\Wc=
    \frac14\int d^4x\,d^4\theta\,du\,\Wc^2\,.
\end{equation}
Here $\Wc$ is the covariant superfield strength
\begin{equation}\label{Wc}
    \Wc=-\frac14(\bar{D}^+)^2 \Vmm\equiv
    \Ac(x_L,\theta^+,\theta^-) +\thetab^+_\alphad\tau^{-\alphad}(x_L,\theta^+,\theta^-)+(\thetabp)^2\tau^{--}(x_L,\theta^+,\theta^-)\,,
\end{equation}
and $\Vmm$ is the non-analytic harmonic connection related to $\V$ by the
harmonic flatness condition
\begin{equation}\label{flateqVmm}
\Dpp V^{--}-\Dmm\V+\left[ \V\,,\, V^{--}\right]_\star = 0\,.
\end{equation}
The whole effect of the considered deformation in the above action comes
from the structure of $\Wc$ due to the presence of the star commutator
in the equation \eqref{flateqVmm} defining $\Vmm$. As a consequence of
the latter, \eqref{Wc} satisfies the condition
\begin{equation}\label{flateqWc}
\Dpp\Wc+\left[ \V\,,\,\Wc\right]_\star = 0\,.
\end{equation}
It is not hard to prove that the only contribution to the entire action
is the superfield $\Ac$ in \eqref{Wc} (see \cite{Ferrara:2004zv}). Thus,
the invariant action is reduced to
\begin{equation}\label{actionAc}
    S=\frac14\int d^4x\,d^4\theta\,du\,\Ac^2\,.
\end{equation}
Once again we refer to \cite{DeCastro:2005vb} for details of the
calculations leading to the relevant components of $\Ac$. Finally the bosonic
limit of the action, after performing the minimal SW map \eqref{SW}, is 
\begin{align}
    S_{bos}=\int\,d^4x\,\Bigl[&-\frac{1}{2}\tilde{\phi}\square\phib
    -\frac12 (b^2\, c^2)^{3/2} \tanh X \partial_{\alpha\alphad}\phib
    \partial^{\alpha\alphad}\phib\square\phib
    +\frac14 \frac{\widetilde{D}^2}{\cosh^2{X}}\nonumber\\
    &-\frac1{16}\widetilde{F}^2\cosh^2X
    +\frac14b^2(c\cdot\widetilde{F})^2\phib^2\frac{\sinh^2 X}{X^2}
    +\frac12\phib(b\cdot\widetilde{D})
    (c\cdot\widetilde{F})\frac{\tanh{X}}X\Bigr].
\end{align}
This action is invariant under the standard abelian gauge
transformations. Turning off the deformation parameters we are left with
the usual bosonic sector of the undeformed action. Performing the further
field redefinition
\begin{equation*}
    d^{ij}=\frac1{\cosh^2 X}\widetilde{D}^{ij}
    +\phib(c\cdot\widetilde{F})b^{ij}\frac{\tanh X}X ,\quad
    \varphi=\frac1{\cosh^2 X}\left[ \tilde\phi
    +(b^2c^2)^{3/2}(\partial\phib)^2\tanh X \right],
\end{equation*}
the bosonic action can be transformed into a simple form
\begin{equation}\label{finalaction}
S_{bos}=\int\,d^4x\; \cosh^2{X} \left[- \frac12\varphi\square\phib
+ \frac14 d^{ij}d_{ij}
-\frac1{16}\tilde{F}^{\alpha\beta}\tilde{F}_{\alpha\beta} \right].
\end{equation}
From this expression it is obvious that we cannot disentangle the
interaction between the gauge field and $\phib$ by any field
redefinition. This is similar to the singlet case \cite{Ferrara:2004zv,
Araki:2004mq}, where a scalar factor $(1 + 4I\bar\phi)^2$ appears
instead of $\cosh^2 X\,$. Note that the bosonic action
\eqref{finalaction} involves only squares $c^2$ and $b^2$, so it
preserves space-time $Spin(4)$= SU(2)$_{\text{L}} \times
$SU(2)$_{\text{R}}$ symmetry and SU(2) R-symmetry as in the singlet
case. This property is similar to what happens in the deformed Euclidean
${\cal N} = (1/2,1/2)$ Wess-Zumino model where the deformation parameter
$C^{\alpha\beta}$ also appears squared \cite{Seiberg:2003yz}.  However,
we know that the fermionic completion of \eqref{finalaction} will
explicitly include both $c^{\alpha\beta}$ and $b^{ik}$ \cite{In prep},
so these two symmetries are broken in the total action.  This feature
also compares with the breaking of Lorentz symmetry in the deformed
${\cal N} = (1/2,1/2)$ gauge theory action, due to fermionic terms
\cite{Seiberg:2003yz}.  Though the string interpretation of the
QS-deformation is known \cite{Ferrara:2004zv}, the possible stringy
origin of the non-singlet case ---e.g. as some special ${\cal N}=4$
superstring background--- is still unclear.

\end{document}